\documentclass[epj]{webofc}
\usepackage[utf8]{inputenc}
\usepackage[varg]{txfonts}   
\usepackage{booktabs}
\usepackage{xcolor}
\definecolor{darkred}{rgb}{0.4,0.0,0.0}
\definecolor{darkgreen}{rgb}{0.0,0.4,0.0}
\definecolor{darkblue}{rgb}{0.0,0.0,0.4}
\usepackage[bookmarks,linktocpage,colorlinks,
    linkcolor = darkred,
    urlcolor  = darkblue,
    citecolor = darkgreen]{hyperref}
\newcommand{\SU}{\mathrm{SU}}
\newcommand{\dd}{{\rm{d}}}

\newcommand{\real}{{\rm Re\,}}
\newcommand{\Tr}{{\rm Tr\,}}

\usepackage{subfig}
\wocname{EPJ Web of Conferences}
\woctitle{Lattice2017}
\begin{document}
%
\selectlanguage{english}
\title{%
The equation of state with non-equilibrium methods
}
\author{%
\firstname{Alessandro} \lastname{Nada}\inst{1}\fnsep\thanks{Speaker, \email{anada@to.infn.it}} \and
\firstname{Michele} \lastname{Caselle}\inst{1,2} \and
\firstname{Marco}  \lastname{Panero}\inst{1}
}
\institute{%
Department of Physics, University of Turin \& INFN, Turin\\
Via Pietro Giuria 1, I-10125 Turin, Italy
\and
Arnold-Regge Center, University of Turin\\
Via Pietro Giuria 1, I-10125 Turin, Italy
}
\abstract{%
Jarzynski's equality provides an elegant and powerful tool to directly compute differences in free energy in Monte Carlo simulations and it can be readily extended to lattice gauge theories to compute a large set of physically interesting observables. 
In this talk we present a novel technique to determine the thermodynamics of strongly-interacting matter based on this relation, which allows for a direct and efficient determination of the pressure using out-of-equilibrium Monte Carlo simulations on the lattice. 
We present results for the equation of state of the $\SU(3)$ Yang-Mills theory in the confined and deconfined phases. 
Finally, we briefly discuss the generalization of this method for theories with fermions, with particular focus on the equation of state of QCD.
}
\maketitle
\section{Introduction}
\label{sec:intro}

The determination of the equation of state of strongly-interacting matter represents a crucial endeavour in theoretical physics, with many applications in different fields such as nuclear physics and cosmology.
It serves as an input for the analysis of thermal systems such as those created in heavy-ion collision experiments or for the study of the early phases of the Universe itself.
From the theoretical side, the most significant and reliable contribution to this effort comes undoubtedly from the lattice formulation of QCD, which provides a tool for first-principles numerical predictions with ever-increasing precision and accuracy.
In the last few years there have been major advancements in the computation of equilibrium thermodynamics for full QCD with $2+1$ (or more) dynamical quark flavors; still, such calculations require an impressive numerical effort and many systematic effects have to be taken into account.
Thus, recently there has been renovated interest in the study of new ways of computing the equation of state in addition to standard techniques such as the integral method~\cite{Engels:1990vr}: among the latest advancements, we mention studies in a moving reference frame~\cite{Giusti:2016iqr} and with the gradient flow~\cite{Kitazawa:2016dsl}.

The purpose of this paper is to present a novel method for the calculation of the pressure in lattice gauge theories exploiting a well-known result by C.~Jarzynski in out-of-equilibrium statistical mechanics.
The so-called Jarzynski's equality~\cite{Jarzynski:1996ne,Jarzynski:1997ef} relates the difference in free energy $\Delta F$ between two equilibrium states with the average of the exponential of the work done on the system of interest during a transformation between the two states: crucially, such transformations in general will not be performed with the system in thermodynamic equilibrium.
Even more importantly, the average of the exponential of the work must be taken on an ensemble of realizations of this transformation.

In Ref.~\cite{Caselle:2016wsw} Jarzynski's equality was succesfully tested in the context of lattice gauge theories in two different benchmark studies: the first concerning the free energy associated to an interface in the $\mathrm{Z}_2$ gauge model, and the second focusing on the pressure in the $\SU(2)$ gauge theory on a small range of temperatures.

In this work, which is a natural prosecution of the work done in Ref.~\cite{Caselle:2016wsw}, we present a preliminary study of the equation of state of the $\SU(3)$ Yang-Mills theory obtained with non-equilibrium methods based on Jarzynski's equality: the theory without quarks represents a perfect testing ground for new techniques, since one can avoid the complications related to dynamical fermionic fields.
At the same time, the seminal work on $\SU(3)$ thermodynamics of Ref.~\cite{Boyd:1996bx} has been improved in recent years by high-precision determinations obtained with different methods~\cite{Borsanyi:2012ve,Giusti:2016iqr}, that give us the possibility to test the reliability of our technique.

\section{Jarzynski's equality}
\label{sec:jarzynski}

In this section we will state Jarzynski's equality precisely, and analyse in detail its practical implementation in Monte~Carlo simulations.
In order to discuss the non-equilibrium work relation, we start from the second law of thermodynamics in the form of the Clausius inequality
\begin{equation}
 \int_A^B \frac{\delta Q}{T} \leq \Delta S = S(B) - S(A)
\end{equation}
where $S$ is the entropy, $T$ is the temperature, and $Q$ is the heat exchanged with the environment during a transformation between macrostates $A$ and $B$; for an isothermal transformation it can be rewritten as
\begin{equation}
\label{second_law_free_energy}
 W \geq \Delta F = F(B) - F(A)
\end{equation}
using the first law ($\Delta E = Q + W$) and the definition of free energy $F = E - ST$, $E$ being the internal energy.
Let us now consider a system whose dynamics is described by a Hamiltonian $H_\lambda$ that depends explicitly on a certain parameter $\lambda$, e.g. the coupling of the model; the corresponding partition function $Z$ will be
$$ Z(T,\lambda) = \int \dd \Gamma e^{-H(\Gamma, \lambda)/T}$$
where $\Gamma$ indicates a microstate of the system and $k_B=1$.
In this framework, we can think the transformation $A\to B$ to be driven by a change in this parameters, i.e. $\lambda_A \to \lambda_B$.
Moreover, we know that for a microscopic system Eq.~\eqref{second_law_free_energy} is valid only statistically, i.e.
\begin{equation}
\label{second_law_true_free_energy}
 \langle W \rangle \geq \Delta F = F(\lambda_B) - F(\lambda_A)
\end{equation}
where the $\langle...\rangle $ from now on will denote an average over all possible realizations of $\lambda_A \to \lambda_B$ transformation, during which the total work $W$ spent to perform the switch in $\lambda$ is measured.
We can now state the non-equilibrium work relation by C.~Jarzynski~\cite{Jarzynski:1996ne,Jarzynski:1997ef}
\begin{equation}
\label{jarzynski_equality_1}
 \left\langle \exp\left( -\frac{W}{T}\right) \right\rangle = \exp \left(-\frac{\Delta F}{T} \right) = \frac{Z(T,\lambda_B)}{Z(T,\lambda_A)}
\end{equation}
that puts in relation the average of the exponential of the work performed in an isothermal transformation $\lambda_A \to \lambda_B$ with the difference in free energy between the initial and final states or, equivalently, the corresponding ratio of partition functions $Z$.
Using Jensen's inequality, i.e. $\left\langle e^x \right\rangle \geq e^{\langle x \rangle}$, valid for a real variable $x$, it is easy to show that Eq.~\eqref{jarzynski_equality_1} is a generalization for microscopic systems of the second law of thermodynamics.

\subsection{The non-equilibrium work relation for Monte~Carlo simulations}
\label{subsec:jarzynski_montecarlo}

Jarzynski's equality has been derived also in the context of stochastic processes (see for example~\cite{Jarzynski:1997ef,Crooks:1998}) and in particular for Markov chains: thus, the implementation for Monte~Carlo simulations is rather straightforward.
However, before using Eq.~\eqref{jarzynski_equality_1} it is crucial to understand what precisely $W$ is and how in practice the non-equilibrium transformation is performed.
Firstly, the transformation has to be discretized into $N$ intervals, so that at each intermediate step the $\lambda$ parameter changes:
$$  \lambda_0 \to \lambda_1 \to \lambda_2 \to... \to \lambda_N$$
where $\lambda_0$ corresponds to the initial macrostate previously denoted as $A$ and $\lambda_N$ to $B$; the non-equilibrium relation does not depend on the specific protocol used to switch $\lambda$.
We will also have the corresponding set of intermediate configurations $[\phi_n]$ of the system
$$  [\phi_0] \to [\phi_1] \to [\phi_2] \to... \to [\phi_N]\;,$$
where, crucially, $[\phi_0]$ must be a thermalized configuration.
The work $W$ is defined quite naturally as the sum of the difference in the Hamiltonian at each step of the Markov process:
\begin{equation}
\label{jarzynski_work_mc}
 W = \sum\limits_{n=0}^{N-1} \left( H_{\lambda_{n+1}}[\phi(t_n)] - H_{\lambda_{n}}[\phi(t_n)] \right).
\end{equation}
Let us analyze how the entire non-equilibrium transformation is implemented in practice during a Monte~Carlo simulation.
These are the steps that must be followed:
\begin{enumerate}
 \item the non-equilibrium work relation requires the system to be at equilibrium at the beginning of each trajectory; 
 \item \label{list:jar2} we switch the parameters from $\lambda_0$ to $\lambda_1$, following the chosen protocol;
 \item \label{list:jar3} we compute the work done on the system to perform this first change of $\lambda$, simply taking the difference of the Hamiltonian
 $$ H_{\lambda_{1}}[\phi_0] - H_{\lambda_{0}}[\phi_0]; $$
 note that the Hamiltonians are evaluated using the same configuration but different values of $\lambda$;
 \item \label{list:jar4} we update the system with the algorithm of choice to the new configuration $[\phi_1]$ keeping $\lambda$ fixed to $\lambda_1$
 $$[\phi_0] \xrightarrow{\lambda_{1}} [\phi_1] \;;$$
 \item we repeat steps~\ref{list:jar2}, \ref{list:jar3} and \ref{list:jar4} until the transformation is completed. At each step $n$, the parameters are changed following the given protocol $ \lambda_n \to \lambda_{n+1}$,
 the work performed on the system is computed as
 $$ H_{\lambda_{n+1}}[\phi_n] - H_{\lambda_{n}}[\phi_n]\,, $$
 and the system is then updated using the new parameter
 $$[\phi_n] \xrightarrow{\lambda_{n+1}} [\phi_{n+1}] ;$$
 \item at the end of each trajectory, the total work defined in Eq.~\eqref{jarzynski_work_mc} is computed;
 \item a new equilibrium configuration $[\phi_0]$ is generated by thermalizing the system again with $\lambda_0$, and a new trajectory can begin.
\end{enumerate}
It is extremely important to stress that one has to perform several independent realizations of the transformation so that the exponential average of Eq.~\eqref{jarzynski_equality_1} provides reliable results.
The interplay between the number of such realizations, denoted as $n_R$, and the number $N$ of intervals in $\lambda$ is crucial to improve the efficiency of this technique.
We conclude this section by noting that the relation can be extended to non-isothermal transformations (see, for example, Ref.~\cite{Chatelain:2007ts}).

\section{The equation of state with Jarzynski's equality}
\label{sec:eos_with_jarzynski}

Following the work of Ref.~\cite{Caselle:2016wsw}, in this section we will review how to compute the pressure using non-equilibrium transformations in finite-temperature lattice simulations.
We start by considering a model with a given partition function $Z(T)$ and a free energy density $f=-T(\ln Z)/V$, defined on an hypercubic lattice $\Lambda$ of sizes $a N_t \times (aN_s)^3$, with $N_t$ and $N_s$ representing the number of lattice sites in the temporal and spatial directions.
The spatial volume corresponds to $V=(aN_s)^3$, while the temperature is defined as usual as $T=1/(aN_t)$.
Our primary observable is the pressure $p$, that in the thermodynamic limit can be written as
\begin{equation}
\label{pressure_1}
 p = -\lim_{V \to \infty} f = \lim_{V \to \infty} \frac{T}{V} \ln{Z}.
\end{equation}
Other relevant thermodynamical quantities are the energy density $\epsilon$
\begin{equation}
 \epsilon = \frac{T^2}{V} \frac{\partial \ln{Z}}{\partial T}
\end{equation}
and the trace of the energy-momentum tensor $\Delta = \epsilon -3p$, which can be conveniently expressed as
\begin{equation}
 \Delta = T^5 \frac{\partial}{\partial T} \left( \frac{p}{T^4} \right).
\end{equation}
The dimensionless ratio $p(T)/T^4$ can be written as
\begin{equation}
\label{pressure_2}
\frac{p(T)}{T^4} = \frac{N_t^3}{N_s^3} \ln{Z(T)}
\end{equation}
and if we compute differences in $p/T^4$ between two temperatures $T$ and $T_0$, then we have
\begin{equation}
\label{pressure_3}
\frac{p(T)}{T^4} - \frac{p(T_0)}{T_0^4} = \frac{N_t^3}{N_s^3} \ln\left(\frac{Z(T)}{Z(T_0)} \right)
\end{equation}
and it is the $Z(T)/Z(T_0)$ ratio that can be computed directly with Jarzynski's equality using non-equilibrium transformations in which the temperature $T$ is varied.
In practice, the temperature is changed throughout each trajectory by changing the lattice spacing $a$, i.e. by tuning the inverse coupling $\beta$ to the desired values.
Indeed, Eq.~\eqref{jarzynski_equality_1} becomes
\begin{equation}
\label{pressure_4}
\frac{p(T)}{T^4} - \frac{p(T_0)}{T_0^4} = \frac{N_t^3}{N_s^3} \ln \left\langle \exp \left[ -\sum\limits_{n=0}^{N-1} \left( S[\beta_{n+1},U(t_n)] - S[\beta_{n},U(t_n)] \right) \right] \right\rangle = \frac{N_t^3}{N_s^3} \ln \left\langle \exp \left[- \Delta S \right] \right\rangle
\end{equation}
where $\Delta S$ is the total change in the Euclidean action and $\langle...\rangle$ indicates the average on a set of $n_R$ non-equilibrium trajectories performed as described in section~\ref{subsec:jarzynski_montecarlo}.
The protocol to change the parameter (in this case the inverse coupling $\beta$) in the transformation is chosen to be linear in the index $n$ of the intermediate steps, so that
\begin{equation}
\label{puregauge_bc_protocol}
 \beta_n = \beta_0 + n\frac{\beta_N-\beta_0}{N} \equiv \beta_0 + n \, \Delta \beta \;;
\end{equation}
of course, in this notation $T_0 \equiv 1/(a(\beta_0) N_t)$ and $T \equiv 1/(a(\beta_N) N_t)$.
$\Delta S$ is the quantity computed for each trajectory and it is the equivalent of the ``work'' defined in Eq.~\eqref{jarzynski_work_mc}, in which the action $S$ has taken the place of $H/T$.
Before being able to compute the physical value of the pressure, we need to take care of the quartic divergence in $a$: a simple way to do so is to compute the same quantity of Eq.~\eqref{pressure_4} but at $T=0$, i.e. on a symmetric lattice $\widetilde{\Lambda}$ of hypervolume $(a\widetilde{N})^4$.
Then we can subtract it from the result of the finite-temperature simulation so that
\begin{equation}
\label{lattice_pressure_Jarzynski}
\frac{p(T)}{T^4} = \frac{p(T_0)}{T_0^4} + \left( \frac{N_t}{N_s} \right)^3 \ln \frac{ \langle \exp \left[ - \Delta S_{N_t \times N_s^3} \right] \rangle }{ \langle \exp \left[ - \Delta S_{\widetilde{N}^4} \right] \rangle^{\gamma} }
\end{equation}
where the exponent $\gamma = \left( N_s^3 \times N_t \right) / \widetilde{N}^4$ is necessary since the lattice hypervolumes at $T=0$ and $T\neq 0$ are in general different.

\section{Numerical results for the \texorpdfstring{$\SU(3)$}{} pure-glue theory}
\label{sec:numerical_results}

For the Euclidean Yang-Mills action of the lattice theory we chose the standard Wilson action~\cite{Wilson:1974sk},
\begin{equation}
\label{wilson_action}
 S_W = -\frac{2}{g^2} \sum_{x \in \Lambda} \sum_{0 \le \mu < \nu \le 3} \real \Tr U_{\mu\nu} (x)
\end{equation}
where $g$ denotes the (bare) lattice coupling and 
\begin{equation}
\label{plaquette}
U_{\mu\nu} (x) = U_\mu (x) U_\nu \left(x+a\hat{\mu}\right) U_{\mu}^\dagger \left(x+a\hat{\nu}\right) U_{\nu}^\dagger (x)
\end{equation}
is the plaquette. 
The partition function of our theory is 
\begin{equation}
 Z(\beta) = \int \prod_{x \in \Lambda} \prod_{\mu = 0}^{3} \dd U_\mu(x) \exp \left( -S_W (U) \right)
\end{equation}
where $\dd U_\mu(x)$ denotes the Haar measure of the $\SU(N)$ variable at the site $x$ and direction $\mu$.

\begin{figure}[thb]
  \centering
  \includegraphics[width=.8\textwidth]{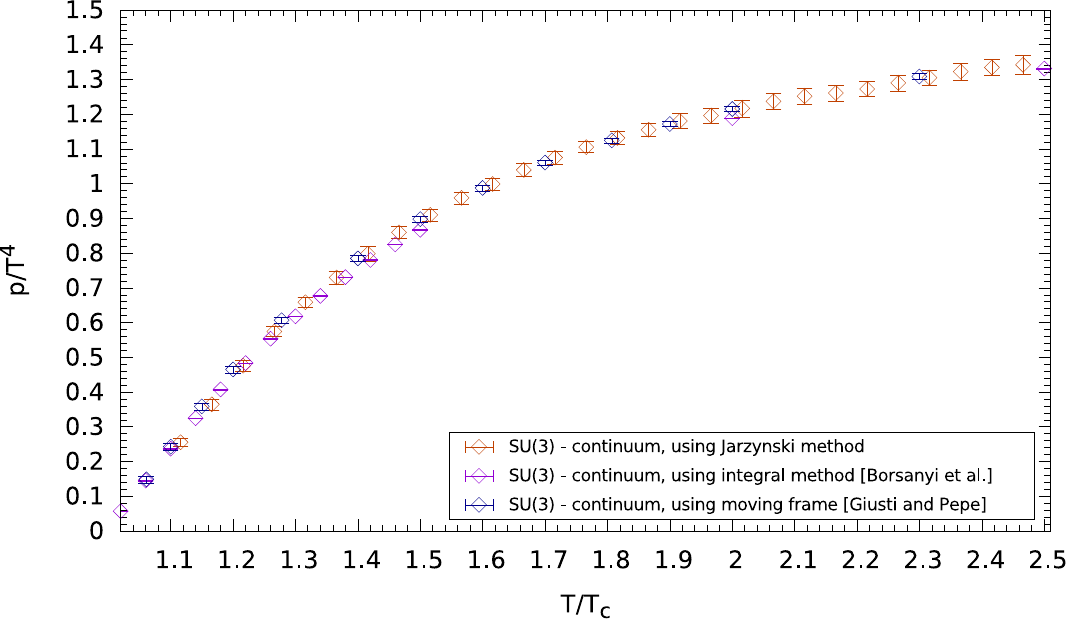}
  \caption{\label{fig:pressure}
  Preliminary continuum-extrapolated results (orange squares) for the pressure in units of $T^4$ in the $[T_c,2.5T_c]$ temperature range obtained with Jarzynski's relation.
  Continuum-extrapolated data from Ref.~\cite{Borsanyi:2012ve} (violet squares) and Ref.~\cite{Giusti:2016iqr} (blue squares) are also presented.}
\end{figure}

The simulations were performed on lattices with $N_t=6,7,8,10$, while keeping the aspect ratio $N_s/N_t > 12$ and in most cases around $16$, in order to avoid finite-size effects. 
The scale is controlled using the values of the Sommer scale $r_0/a$ determined in~\cite{Necco:2001xg}; moreover we used for the critical temperature the value $T_c r_0=0.7457(45)$ computed in~\cite{Francis:2015lha}.
The finite lattice spacing results were first interpolated with spline functions and then the continuum extrapolation was performed with a linear fit in $(1/N_t)^2$. 
Preliminary results for the pressure are illustrated in Fig.~\ref{fig:pressure}.

\begin{figure}[thb]
  \centering
  \includegraphics[width=.8\textwidth]{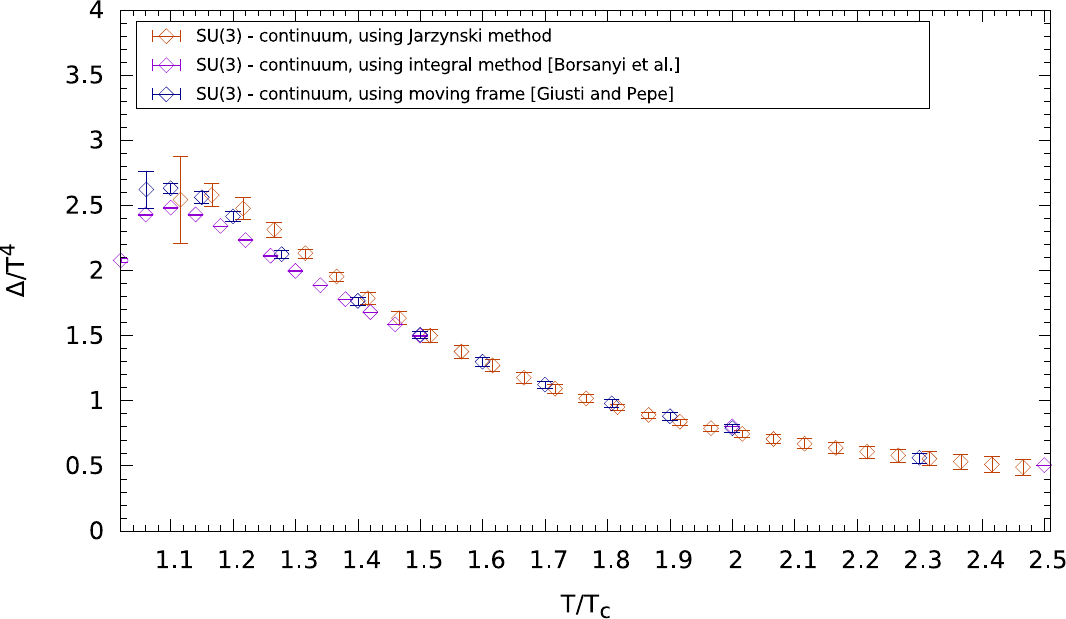}
  \caption{\label{fig:trace}
  Same as Fig.~\ref{fig:pressure}, but for the trace of the energy-momentum tensor $\Delta$ in units of $T^4$.}
\end{figure}

\begin{figure}[thb]
  \centering
  \includegraphics[width=.8\textwidth]{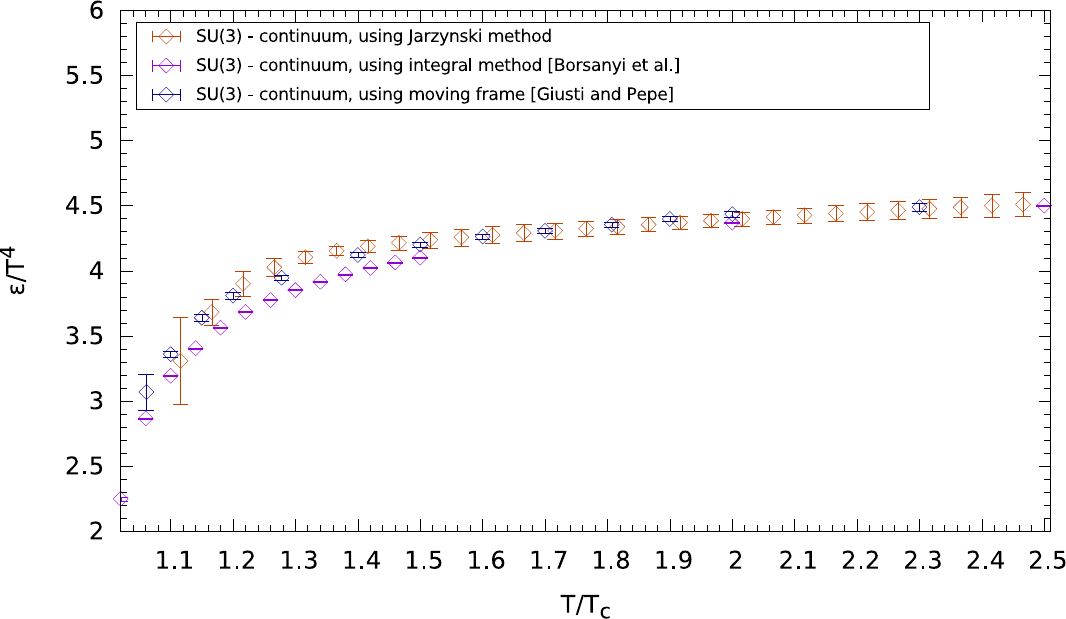}
  \caption{\label{fig:energy}
  Same as Fig.~\ref{fig:pressure}, but for the energy density $\epsilon$ in units of $T^4$.}
\end{figure}

As it can be seen, the data for the pressure obtained using the technique based on Jarzynski's equality are in good agreement with previous high-precision determinations by the Wuppertal-Budapest collaboration~\cite{Borsanyi:2012ve} and more recently by L.~Giusti and M.~Pepe~\cite{Giusti:2016iqr}.
We remark however that these two last computations showed a small but clearly visible discrepancy in the deconfining phase, in particular in the region between $T_c$ and $1.5T_c$; this disagreement becomes gradually smaller as $T$ increases and disappears for $T>3T_c$.
We will not attempt a discussion of the possible origin of this problem here: we limit ourselves to note that our preliminary data for $p/T^4$ seem to agree well with those of Ref.~\cite{Giusti:2016iqr} in the aforementioned region.
In order to investigate this issue more in detail, we also present preliminary data for the trace anomaly $\Delta$ in Fig.~\ref{fig:trace} and the energy density $\epsilon$ in Fig.~\ref{fig:energy}.

First, we note that in order to obtain $\Delta$ it is necessary to numerically derive the pressure $p$ with respect to $T$: this is to be contrasted with the integral method~\cite{Engels:1990vr} and computations in a moving frame, in which a secondary observable such as the pressure requires numerical integration from the quantity that is directly measured on the lattice, i.e. the trace anomaly $\Delta$ or the entropy density $s=(\epsilon+p)/T$.
In order to do so we performed first a Pad\'e interpolation and then the derivation of the resulting fit: preliminary results in Fig.~\ref{fig:trace} and Fig.~\ref{fig:energy} confirm that our method seems to favor the results of Ref.~\cite{Giusti:2016iqr}.
We report that the most delicate temperature region to probe is the one immediately above $T_c$, where a strong dependence on the lattice spacing was observed, leading to larger uncertaintes in the $a\to0$ extrapolation.

As previously discussed in Ref.~\cite{Caselle:2016wsw}, the exponential average of Eqs.~\eqref{jarzynski_equality_1} and \eqref{lattice_pressure_Jarzynski} requires a sample of $n_R$ trajectories that is large enough, in order to converge to the correct result.
The correct way to ensure this is to repeat the transformation in the reverse direction and to check if the results agree with the ``direct'' one; this has been performed in this work, and whenever the agreement was not satisfactory, the transformation was repeated with a new combination of $N$ intermediate steps and $n_R$ realizations.

A possible issue related to this novel technique concerns the use of non-equilibrium transformations that cross the deconfinement phase transition, as in such cases Jarzynski's equality cannot be used.
This fact is confirmed by the strong disagreement between results of trajectories going from the confined to the deconfined phase, and trajectories which perform the same transformation in the reverse direction.
In order to avoid this issue, our non-equilibrium transformations never crossed $T_c$.

\section{Conclusions}
\label{sec:conclusions}

In this work a new determination of the $\SU(3)$ equation of state has been performed, using a technique based on Jarzynski's equality: preliminary results obtained with out-of-equilibrium Monte~Carlo simulations show good agreement with past determinations.
The new method proved to be very efficient, since only the starting configuration has to be at equilibrium, but also highly reliable, as each transformation has to be in agreement with the same transformation performed in the opposite direction.
Jarzynski's equality has a wide range of possible applications in lattice gauge theory, as the problem of computing differences in free energy is a very general and common one; however the most natural prosecution of this work is the application of this technique to the equation of state in full QCD.
A particularly intriguing idea is to set up non-equilibrium transformations in which both the temperature $T$ (via the inverse coupling $\beta$) and the bare masses of the fermions are changed simultaneously.

\begin{acknowledgement}
 The simulations were run on GALILEO and MARCONI supercomputers at CINECA. We thank M.~Hasenbusch and R.~Sommer for helpful comments and insightful discussions.
\end{acknowledgement}
\bibliography{lattice2017}

\end{document}